\documentstyle[art10,fleqn]{article}  
\begin{document}
\baselineskip=12pt
\setcounter{page}{1}
\def\NL{\hfill\break}
\def\NI{\noindent}
\def\SL{{\vskip 0.15in}}
\def\NP{\vfil\eject}
\long\def\UN#1{$\underline{{\vphantom{\hbox{#1}}}\smash{\hbox{#1}}}$}
\def\today{\ifcase\month\or January\or February\or March\or April\or
 May\or June\or July\or August\or September\or October\or November\or
 December\fi \space\number\day, \number\year}
 \catcode`@=11
\def\@cite#1{[#1]}
\def\@refe#1{#1}
\def\@biblabel#1{{\normalsize\bf{#1}}}
\def\refe{\@ifnextchar [{\@tempswatrue\@citexr}{\@tempswafalse\@citexr[]}}
\def\@citexr[#1]#2{\if@filesw\immediate\write\@auxout{\string\citation{#2}}\fi
  \def\@citea{}\@refe{\@for\@citeb:=#2\do
    {\@citea\def\@citea{,}\@ifundefined
       {b@\@citeb}{{\bf ?}\@warning
       {Citation `\@citeb' on page \thepage \space undefined}}%
\hbox{\csname b@\@citeb\endcsname}}}{#1}}
 \catcode`@=12
\noindent
{\large\bf
Irreversible Multilayer Adsorption
}

\vspace{1cm}

\noindent
P. Nielaba, V. Privman$^1$, J.-S. Wang$^2$

\noindent
Institut f\"ur Physik, Staudingerweg 7, Universit\"at Mainz, 6500 Mainz, FRG

\noindent
1) Physics Department, Clarkson University, Potsdam, NY 13699-5820, USA

\noindent
2) Physics Department, Hong Kong Baptist College, 244 Waterloo Road,
Kowloon, Hong Kong

\centerline{ }

\centerline{ }

\noindent
{\bf Abstract.}
Random sequential adsorption (RSA) models have been studied [1]
due to their relevance to deposition processes on
surfaces. The depositing particles are represented
by hard-core extended objects; they are not allowed to overlap.
Numerical Monte Carlo studies and analytical considerations
are reported for 1D and 2D models of multilayer adsorption processes.
Deposition without screening is
investigated,
in certain models the density
may actually increase away from the substrate.
Analytical studies of the late stage coverage behavior show the
crossover from exponential time dependence for the lattice case to the
power law behavior in the continuum deposition. 2D lattice and continuum
simulations rule out some ``exact'' conjectures for the jamming coverage.
For the deposition of dimers on a 1D lattice
with diffusional relaxation
we find that
the limiting
coverage $(100 \%)$ is approached according to the
$\sim 1/\sqrt{t}$ power-law preceded, for fast diffusion,
by the mean-field crossover
regime with the intermediate $\sim 1/t$ behavior.
In case of $k$-mer deposition ($k>3$) with diffusion the void
fraction decreases
according to
the power-law $t^{-1/(k-1)}$.
In the case of RSA of lattice hard squares in 2D with diffusional relaxation
the approach to the full coverage is
$\sim t^{-1/2}$.

\centerline{ }

\centerline{ }

\noindent
{\large\bf 1. Kinetics of multilayer adsorption \hfil}

\centerline{ }

\noindent
In this section
we study [1] the
deposition of $k$-mers of length $k$ on the linear periodic 1D
lattice of spacing $1$, and the deposition of square-shaped $(k \times
k)$-mers on
the periodic square lattice of unit spacing in 2D. The deposition
site is chosen at random.
The time scale, $t$,
is fixed by having exactly $L$ deposition attempts
per unit time.
We studied systems of sizes up to $L=10^5$ in  1D
and $L \times L =1000^2$ in 2D.
Various Monte Carlo runs
(for $k=2,3,4,5,10$ in 1D, and
$k=2,4$ in 2D)
went
up to $t=150$.
Our numerical values are consistent with exact results for layer $n=1$ in 1D
[2].

We first report on a model with decreasing layer coverage.
Only if {\it all\/} the lattice segments in the selected
landing site are already covered by exactly $(n-1)$ layers, the
arriving $k$-mer is deposited, increasing the coverage to $n\;$
($n \geq 1$).
For small $t$, the coverage (fraction of the total volume covered by
depositing particles)
increases according to $\theta_{n}(t)
\propto t^{n}$, as expected from the
mean-field theory.
Our results suggest that for lattice models
the fraction of the
occupied area in the $n$th layer, $\theta_n (t)$, approaches
the saturation value exponentially,
$ \theta_n (t) \approx
 \theta_n (\infty ) + B_n {\exp}{(-t / \tau_n)}$.
However, the
jammed state in
the higher layers {\it in the deposition without overhangs\/}
contains more gaps the
larger is the $n$ value. The growth  in the higher
layers proceeds more and more via uncorrelated
``towers".
We find
that the jamming coverages vary according to a power law, with no length
scale, reminiscent of critical phenomena,
\begin{equation}
 \theta_n ( \infty ) - \theta_\infty ( \infty ) \approx
 {A \over n^\phi }
\end{equation}
Within the limits
of the numerical accuracy the
values of the exponent $\phi$ are universal
for $k \geq 2 $.
Based on numerical data analysis we find the estimates
$ \phi({\rm 1D}) = 0.58 \pm 0.08$ and
$  \phi({\rm 2D}) = 0.48 \pm 0.06$. These values are most likely exactly
$1/2$ as suggested by analytical random-walk arguments [3].

Now we report on a model with decreasing layer coverage.
For layers $n>1$ the deposition
is successful only if no gaps of size $k$ or larger are covered.
Thus, the deposition is always allowed if all the ``supporting'' $k$ sites in
the $(n-1)$st
layer are filled or have only small internal gaps.
The coverage {\sl at short times\/}
decreases with layer number.
However, for the particular deposition
rule considered here the coverage in layer $n$ eventually exceeds that in layer
$(n-1)$ at larger times.
This unexpected behavior was found numerically for all layers $n \leq 55$ and
for
all $k$ ($k=2,3,4,5,10$) studied.
We found clear evidence of the power-law behavior (eq.(1)), with $A<0$.
We found the power $\phi$
to be universal for
all $k$ studied, $
 \phi = 0.3 \pm 0.15 $.
When large enough covered
(by $k$-mers or gaps of sizes up to $k-1$) regions
have formed in layer $(n-1)$, then the deposition with overhangs beyond those
regions will be delayed. Thus, there will be some
preference for higher density in layer $n$ especially near the ends of the
regions occupied in layer $(n-1)$.
To test the above suggestion, we
considered the following monolayer dimer-deposition model.
We select randomly $\rho L/2$ dimers
and make the $\rho L$ sites thus selected unavailable for deposition for
times $0 \leq T \leq T_s$. A ``sleeping time'' $T_s$
for fraction $\rho$ of lattice sites (grouped in dimers) in monolayer
deposition supposedly will
model effect of disallowed overhangs over gaps of size larger than 1
in the lower layer on the multilayer deposition
in layer $n$ provided we loosely identify $T_s \propto n$.
Indeed, our multilayer data
suggest that times needed to build up the $n$th layer coverage
grow linearly with $n$. For instance times $T_{1/2}$ defined via
$ \theta \left( T_{1/2} \right) = {1 \over 2} \theta_n (\infty )$,
grow according to
$ T_{1/2} \simeq \tau n $,
where the coefficient $\tau$ is of order 1.
After time $T_s$ all the blocked sites are released and can be occupied in
subsequent deposition attempts.
We find that the variation of the
jamming coverage $\propto T_s^{-\phi}$,\ $\phi \simeq 1/3$.

\noindent
{\large\bf 2. Continuum limit in RSA}

\centerline{ }

\noindent
We consider [1] the deposition of objects of
size $l$ on a $1$D substrate of size $L$.
$R$ is the rate of
random deposition attempts per unit
time and volume.
The lattice approximation is introduced by choosing the cubic mesh size
$b = l / k$.
The lattice
deposition is defined by requiring
that the objects of size $l$ can only deposit in sites consisting of
$k$ lattice units.
The late stage of the deposition (after time $\tau$) in continuum
can be described as filling up of
voids small enough to accommodate only one
depositing object.
At this time
$\tau$, the density
of those small gaps (of various sizes) will be $\rho$.
For lattice models, a similar picture
applies for
$k >> R l \tau $.
Typical
small gaps can be assumed [4]
of sizes $k+n \ (n=0,1,\ldots ,k-1)$
with density $\rho / k$ at
time $\tau$, and will be filled
up at the rate $R b (n+1)$
per unit time. We will consider
$t>> \tau$ so that no new small gaps are
created by the filling up of large gaps.
Then the density $\Omega$ of each type of the small
gaps
will have the time dependence
$ \Omega (n) = (\rho / k) \exp (-R b (n+1)
(t - \tau )) $.
In each deposition event with rate
$R b \Omega ( n) (n +1)$ per unit time
the coverage is increased by $(l/L)$.
Thus, we have
$ {d \theta / dt} \simeq \sum_{n=0}^{k-1}
{(R b l \rho / k )}
(n+1)  \exp \{ - R b
(n+1) (t- \tau ) \} $.
After integration we
get the asymptotic ($t >> \tau$) estimate,
generalized to $D$ dimensions [1]:
\begin{equation}
 \theta_k (t) = \theta_k (\infty ) - {\rho l^D \over k^D}
\sum\limits_{n_1=0}^{k-1} \ldots
\sum\limits_{n_D=0}^{k-1}
\exp \left\{ - \left( R l^D t \over k^D \right)
\prod_{m=1}^D (n_m+1)  \right\} \quad
\end{equation}

\noindent
We consider some special limits.
For $k$
fixed, the ``lattice" long time behavior
sets in for $R l^D t >> k^D$. In this limit
the $n_j = 0$ term in the sums in eq.(2) dominates:
$ \theta (t) \approx \theta (\infty ) - { (\rho l^D / k^D )}
{ \exp}{(- R l^D t / k^D) } $.
\NI Thus, the time decay constant increases as $k^D$,
The continuum limit of eq.(2) is obtained for $k^D >> R l^D t$.
In this limit one can convert the sums to integrals.
Recall that all the expressions here apply only for $t >> \tau$
and $k^D >> R l^D \tau$, where $R l^D \tau$ is a fixed quantity of
order 1. Thus, the large-$k$ and large-$t$ conditions are simply
$k >> 1$ and $t >> 1/ ( R l^D )$.
The latter condition allows us to evaluate the integrals
asymptotically, to the leading order for large $t$, which yields
$ \theta (t) \approx \theta (\infty ) - {( \rho /
(D-1)! \; R t )}
 \left[ \ln \left(
R l^D t \right) \right]^{D-1} $.
\NI The asymptotic $ (\ln t) ^{D-1}  t^{-1} $ law was
derived in Ref. [4] for the continuum deposition of cubic objects.

\noindent
We evaluated  $\theta_k
(\infty )$ numerically for system sizes $L/k=200$.
Our data suggest a fit of the form
$ \theta_k(\infty) = \theta_{\infty}(\infty) +
        {(A_1 / k)} + {(A_2 / k^2)} + \ldots $.
By
standard manipulations to cancel
the leading $1/k$ term, followed by a further extrapolation to
$k \to \infty$, we arrived at the estimate
$\theta_{\infty}(\infty) = 0.5620 \pm 0.0002$.
The errors are small enough to
rule out the conjecture of Pal\'asti [5],
and its generalization for finite $k$,
which state that the jamming coverages for the $2D$
$\, (k \times k)$ oriented squares
are equal to the {\it squared\/} jamming
coverages of the corresponding $1D$ $k$-mer models.  The latter are
known exactly [2].



\noindent
{\large\bf 3. Effects of diffusional relaxation}

\centerline{ }

\noindent
In this section we report
numerical studies [1] of the effects of diffusion on
RSA in 1D and 2D.
In the deposition
of $k$-mers on a 1D linear lattice, holes of $k-1$ sites
or less cannot be reached by deposition. Diffusion of the
deposited objects can combine small holes to form larger landing sites
accessible to further deposition attempts leading
to a fully covered lattice at large times.
For large times the holes are predominantly single-site
vacancies which hop due to $k$-mer diffusion. They must be brought together
in groups of $k$ to be covered by a depositing $k$-mer. If the
deposition rate is small, the $k$-site
holes may be broken again by diffusion before a successful deposition attempt.
Thus the process of $k$-mer deposition with diffusion will reach its
asymptotic large-time behavior when most of the empty space is in
single-site vacancies. The approach of the coverage to  $1$
for large times will then be related
to the reaction
$k {\cal A} \to inert$
with partial reaction probability on each encounter of $k$ diffusing
particles ${\cal A}$.
Scaling arguments indicate
that the
particle density for $k \geq 3$
will follow the mean-field law $\propto t^{-1/(k-1)}$
for large times, with possible
logarithmic corrections for $k=3$
(borderline).
This corresponds to
$1-\theta(t\hbox{-large}) = t^{-1/(k-1)}$
in deposition.
Now we study the effect of diffusional relaxation
in 1D dimer deposition.
At each Monte Carlo step a pair of adjacent sites
on a linear lattice ($L=2000$) is chosen at random.
Deposition is attempted with
probability $p$ or diffusion otherwise with equal
probability to move one lattice spacing to the left or right.
The time step $\Delta T =1$ corresponds
to $L$ deposition/diffusion-attempt Monte Carlo steps.
We define the time variable $t = pT$.
Our Monte Carlo results were obtained for $p=0.9, 0.8,
0.5, 0.2$.
The coverage
increases monotonically with $(1-p)/p$ at
fixed $pT$.
For $p<1$ we obtain $\theta(\infty)=1$, for $p=1$,
$\theta (T) = 1 - \exp \left\{ -2 \left[ 1 - \exp (-T) \right]
\right\} \to
 1 - {\sl e}^{-2
}  < 1 .$
The convergence of ($1-\theta$)
to the limiting value at $t=pT=\infty$ is exponential
without diffusion. Small diffusional rates lead to the asymptotically
$\sim t^{-1/2}$ convergence to $\theta (p<1,
t=\infty) =1 $. For faster diffusion, the onset of the limiting
behavior is preceded by the region of $\sim t^{-1}$ behavior followed
by a crossover to $\sim t^{-1/2}$ for larger times.
In the cases $k=3,4$
the large time results are roughly consistent with the mean-field relation.
For all $p$ values studied the void area is
dominated by the single vacancies precisely in the regime where the
mean-field law sets in.
For $k = 2$ the single-site vacancies take over for
$t^>_\sim 2$. For fast diffusion there follows
a long crossover region from the initially mean-field to the
asymptotically fluctuation behavior.

Now we report on collective effects in RSA of diffusing hard squares.
In each Monte Carlo trial of our simulation on a $L\times L$ square
lattice, a site is chosen at
random with deposition probability $p$.
Only if the chosen site and its four
nearest-neighbor sites are all empty the
deposition is performed.
A diffusion move by one lattice spacing
is made if the targeted new site and its nearest neighbors are all
empty.
Numerical estimates were obtained for
the coverage and
the ``susceptibility''
$\chi = L^2 \bigl[ \langle m^2 \rangle -
\langle |m| \rangle^2 \bigr] $,
where the average $\langle\ \rangle$ is over independent runs.
The order parameter was defined
by assigning ``spin'' values $+1$ to particles on one of the
sublattices and $-1$ on another sublattice.
The effective domain size, $\ell (T)$, was defined
by $\ell  = 2L \sqrt{\langle m^2\rangle}$.
For $p=1$,
the approach of $\theta(T)$ to the jamming coverage $\theta (\infty) \simeq
0.728 < 1$\ is
exponentially fast.  With
diffusion, one can always reach the full coverage
$\theta( \infty ) = 1$.  However, the approach to the full
coverage is slow, power-law.
Here the coverage growth mechanism for large times
is due to interfacial dynamics.
The void space at late times consists of
domain walls separating spin-up and spin-down ordered regions.
Since a typical domain has area $\sim \ell^2 (T)$ and boundary
$\sim \ell(T)$, we anticipate that for large times
$ 1 - \theta (T) \propto \ell^{-1} (T)$.
We found that the data roughly fit the power law,
$ 1- \theta(T) \propto T^{-1/2}$, for $T > 10^3$.
Thus, the RSA quantity $1-\theta(T)$ behaves analogously to
the energy excess in equilibrium domain growth problems.
The ``susceptibility'' $\chi$ for a given finite size $L$ has a
peak and then decreases to zero, indicating long-range order for
large $T$.  The peak location seems size-dependent, at
$T_{\rm peak} \propto L^2$.
Since finite-size effects set in for $\ell (T)
\sim L$, which given the ``bulk'' power law $\ell (T) \sim T^{1/2}$
leads precisely to the criterion $T \sim L^2$, we expect this
maximum in fluctuations to be a manifestation of the ordering
process at high densities.

\vspace{1cm}

\noindent {\large\bf Acknowledgments:}
The authors wish to thank Prof. Kurt
Binder for helpful discussions,
and to acknowledge the sponsorship
of the
SFB 262 of the DFG.
P.N. thanks the DFG for a Heisenberg fellowship.

\centerline{ }

\centerline{ }

\noindent
{\large\bf References:}

\vspace{0.34cm}

\noindent
[1] P. Nielaba, V. Privman and J.--S. Wang, J. Phys. {\bf A\ 23},
L1187 (1991);
P.~Nielaba and V.~Privman, Mod. Phys. Lett. {\bf B 6},
533 (1992);
V.~Privman and P.~Nielaba, Europhys. Lett. {\bf 18}, 673
(1992);
V.~Privman, J.--S.~Wang and P.~Nielaba, Phys. Rev. {\bf B1
43}, 3366 (1991); J.--S. Wang, P. Nielaba and V. Privman, Mod.
Phys. Lett. {\bf B} (in press) and references therein.

\vspace{0.34cm}

\noindent
[2] J.J. Gonzalez, P.C. Hemmer and J.S. H{\rlap /}oye,
Chem. Phys. {\bf 3}, 228 (1974).

\vspace{0.34cm}

\noindent
[3] R.~Hilfer and J.--S.~Wang, J.~Phys. {\bf A~24}, L389 (1991);
V.~Privman and J.--S.~Wang, Phys.~Rev.~{\bf A~45}, R2155 (1992).

\vspace{0.33cm}

\noindent
[4] R.H. Swendsen, Phys. Rev. {\bf A 24}, 504 (1981).

\vspace{0.33cm}

\noindent
[5] I. Pal\'asti,  Publ. Math. Inst. Hung. Acad. Sci.






















\vfill
\eject

\end{document}